\documentclass[aps,prd,twocolumn,balancelastpage,amsmath,amssymb,nofootinbib,10pt,showpacs]{revtex4}

\newcommand{\degrees}{\ensuremath{^\circ}}

\usepackage{color}         
\usepackage{graphics}      
\usepackage[pdftex]{graphicx}      
\usepackage{verbatim}
\usepackage{subfigure}
\usepackage{bm}            
\usepackage{amsmath}
\usepackage[colorlinks=false]{hyperref}  
\usepackage[abs]{overpic}

\begin{document}

\title{QED Radiative Corrections to Low-Energy M\o ller and Bhabha Scattering}
\author         {Charles S. Epstein}
\email          {cepstein@mit.edu}
\author{Richard G. Milner}
\affiliation{Laboratory for Nuclear Science, Massachusetts Institute of Technology, Cambridge, MA 02139 USA}

\date{\today}


\begin{abstract}

We present a treatment of the next-to-leading-order radiative corrections to unpolarized M\o ller and Bhabha scattering without resorting to ultra-relativistic approximations.  We extend existing soft-photon radiative corrections with new hard-photon bremsstrahlung calculations so that the effect of photon emission is taken into account for any photon energy.  This formulation is intended for application in the OLYMPUS experiment and the upcoming DarkLight experiment, but is applicable to a broad range of experiments at energies where QED is a sufficient description.

\end{abstract}

\maketitle

\section{Motivation}

With the development of new precision physics experiments on the Intensity Frontier using lepton beams on targets containing atomic electrons, interest has been renewed in M\o ller and Bhabha scattering as important signal, background, and luminosity-monitoring processes.  Two such experiments are the subject of current attention at the MIT Laboratory for Nuclear Science: DarkLight \cite{darklight} and OLYMPUS \cite{olympus}.  These experiments require calculations of the M\o ller and Bhabha processes including next-to-leading-order radiative effects.

The DarkLight experiment aims to search for a massive dark-sector boson by precisely measuring the process $e^- p \rightarrow e^- p e^+ e^-$.  It will use the 100 MeV electron beam at the Jefferson Lab Low Energy Recirculator Facility incident on a gaseous hydrogen target.  DarkLight aims to measure all four final-state particles in a fourfold coincidence.  At the design luminosity of $\sim$$10^{36}$ $\text{cm}^{-2}\text{s}^{-1}$ and at such low energies, M\o ller electrons and associated radiated photons induce an enormous background of secondary particles.  Careful study is necessary to understand and minimize the backgrounds masking the comparatively rare signal process.

The OLYMPUS experiment aims to measure the ratio of positron-proton to electron-proton elastic scattering cross-sections in the effort to quantify the contribution of two-photon exchange.  OLYMPUS acquired data with 2 GeV alternating electron and positron beams incident on a hydrogen target \cite{target} at the DORIS storage ring at DESY.  M\o ller/Bhabha calorimeters placed at the symmetric angle (90\degrees$_\text{CM}$=1.29\degrees$_\text{lab}$) were used as one of the luminosity monitors.   Precise luminosity monitoring is important to normalize the separate electron and positron datasets and form the cross-section ratio.  Since electron-electron and positron-electron scattering are the only processes in the experiment that can be fully described by QED, they are the most suitable choices for normalization.  As a result, knowledge of their cross-sections including radiative corrections is essential to forming the final result.

A Monte Carlo approach has been identified as the preferred method of treating the radiative corrections for both of these experiments.  This approach stands in contrast with traditional soft-photon radiative corrections, which are typically included as a multiplicative factor to the Born cross section:
\begin{equation}
\dfrac{\mathrm{d}\sigma}{\mathrm{d}\Omega}{\bigg \vert}_\text{soft} = (1+\delta) \frac{\mathrm{d}\sigma}{\mathrm{d}\Omega}{\bigg \vert}_\text{Born}
\label{standardRC}
\end{equation}

\noindent with $\delta = \delta(\Delta E, \Omega)$.  This traditional method requires defining a cut-off $\Delta E$: the maximum amount of energy a photon can carry away for which the event passes acceptance cuts.  For an experiment having spectrometers with small, well-defined energy and angular acceptances, this formulation of the radiative corrections can be applied easily.  However, for experiments with irregular acceptances, energy resolutions that may have a complex dependence on angle, or coincidence measurements, it is not feasible to quantify the radiative corrections solely by $\Omega$ and $\Delta E$.  An effective way to convolve the effects of radiation with these constraints is to perform Monte Carlo simulation.  There have already been Monte Carlo implementations of the radiative corrections such as MERADGEN (for M\o ller) \cite{meradgen} and BabaYaga@NLO (for Bhabha) \cite{babayaga1,babayaga2}, but the two use different formalisms and we require a consistent treatment.  Further, neither of these are flexible enough to meet the needs of OLYMPUS or DarkLight.

\begin{figure}
\centering
\includegraphics[width=0.475\textwidth]{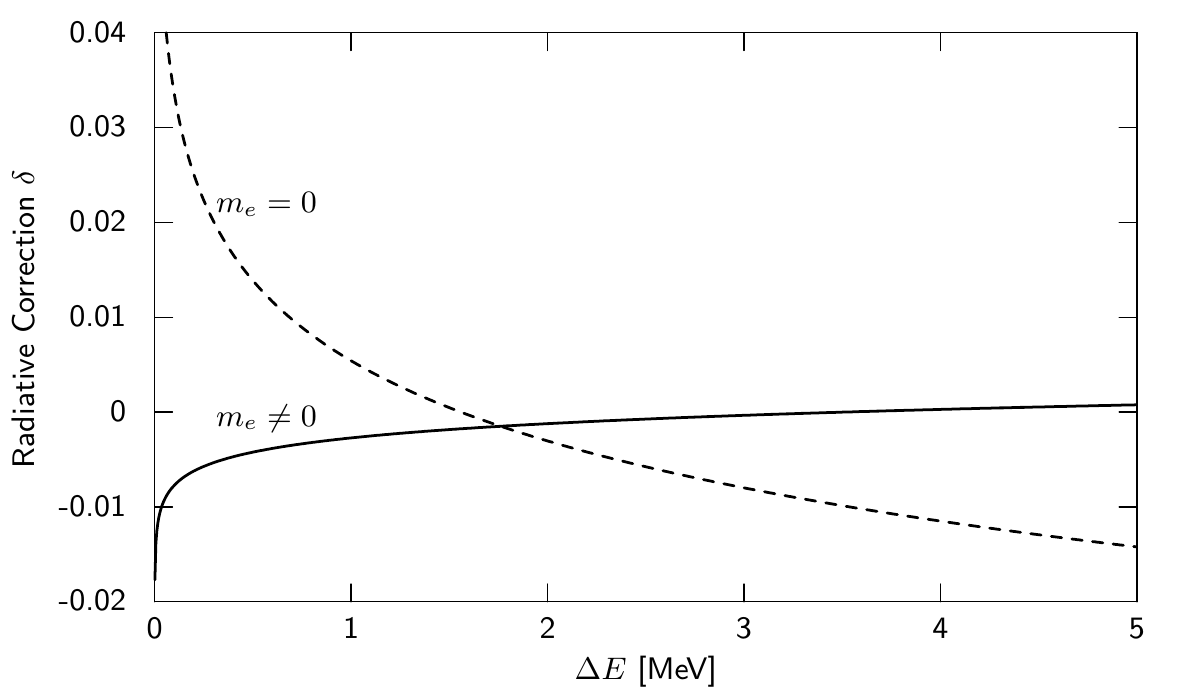}
\caption{Comparison of the M\o ller radiative correction term, $\delta$, for a 100 MeV DarkLight beam at 5$^\circ$ in the CM frame.  With $m_e = 0$ (Tsai \cite{tsai}), the downward-sloping behavior is unphysical; this is fixed when the electron mass is taken into account (Kaiser \cite{softcorrections}).}
\label{DE_Comparison}
\end{figure}

Previous radiative corrections to M\o ller and Bhabha scattering in the traditional approach \cite{tsai,arbuzov,mera,denner} have often made use of ultra-relativistic approximations in which the electron mass is assumed to be negligible relative to the momentum transfer.  While this is sufficient for OLYMPUS where $Q^2_\mathrm{sym}$$\sim$$10^3\ \mathrm{(MeV/c)}^2$ at the symmetric angle, it will not work for DarkLight.  For the majority of the scattering solid-angle, the approximation $m_e^2 \ll Q^2$ does not hold, and the traditional soft-photon radiative corrections exhibit not only inaccurate, but unphysical behavior.  Figure \ref{DE_Comparison} illustrates this: a proper radiative correction factor $\delta(\Delta E, \Omega)$ should decrease as $\Delta E$ decreases, indicating the obvious conclusion that fewer events are expected in a smaller energy window.  However, when the electron mass is neglected in a region where it is important, this behavior flips: the radiative corrections \emph{increase} with decreasing $\Delta E$.  This is unphysical and is one of the primary motivations for this work, which is required if we are to have any reliable analysis at DarkLight-scale energies.  In particular, for DarkLight, $m_e^2/t > 0.1$ outside the lab-frame region of 0.93$^\circ$-31.98$^\circ$, and the flip occurs at approximately 10$^\circ$ in the center-of-mass (CM) frame, which excludes the area outside approximately 0.5$^\circ$-49$^\circ$ in the lab frame.  Since for DarkLight we are interested in electrons at both very small and large angles, it is clearly crucial to include the electron mass.  Nearly all existing formulations were intended for high energy scattering (e.g. \cite{tsai,denner}), and only recently has there been attention to including the electron mass.  

In a paper by N. Kaiser \cite{softcorrections}, the radiative corrections for soft-photon emission in both M\o ller and Bhabha scattering were performed in a consistent approach and without ultra-relativistic approximations.  There has also been an additional recent treatment of the radiative corrections to M\o ller scattering beyond the ultra-relativistic limit \cite{pradcorrections}; however, we do not use it as there is no matching formulation for Bhabha scattering.  In this work, we have extended the results of Kaiser with exact single hard-photon bremsstrahlung calculations.  Since the energies of interest are quite low, only QED interactions have been included.  The calculations, containing no ultra-relativistic approximations, permit a complete analysis of the next-to-leading-order radiative corrections for both M\o ller and Bhabha scattering in the low energy regions of interest.  The results have been packaged in the form of a new C++ Monte Carlo event generator, which will be described in a future publication.

Notably, the scattering of low-energy positrons off atomic electrons allows an additional final-state: annihilation to two or more photons.  This process is important to OLYMPUS since the M\o ller/Bhabha calorimeters cannot distinguish electrons, positrons, and photons.  An additional paper will describe the efforts of our group to characterize the pair annihilation process in the same approach as we have done here for M\o ller and Bhabha scattering.

\section{Treatment of the Radiative Corrections}

\begin{figure}
\centering
\includegraphics[width=0.475\textwidth]{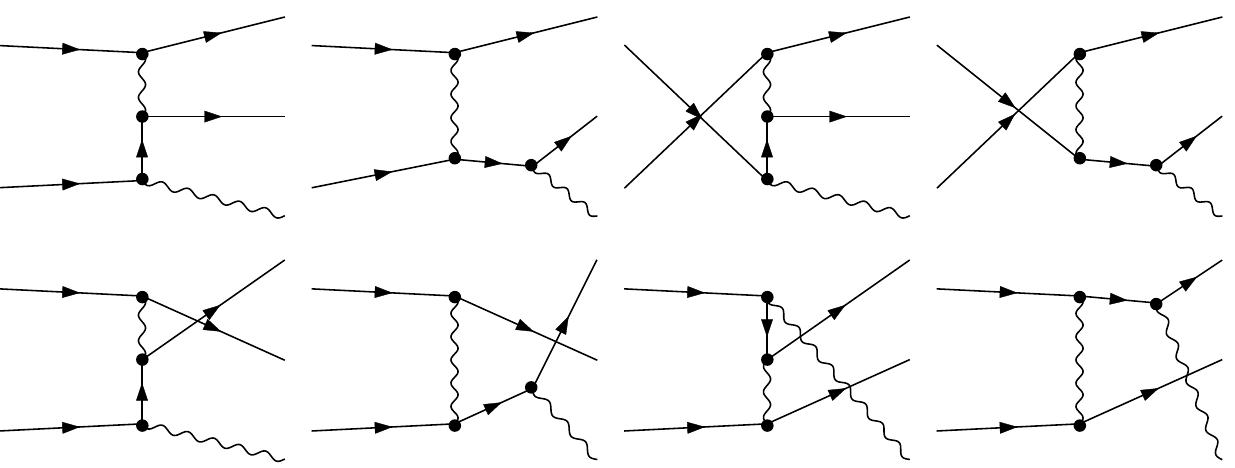}
\caption{Feynman diagrams for radiative M\o ller scattering}
\label{mollerdiagrams}
\end{figure}

Our treatment of the radiative corrections is to divide the events into two categories corresponding to the emission of photons with energy above or below a cutoff, $\Delta E$, that divides the ``soft'' and ``hard'' regimes.  In the soft regime, the events are described by elastic electron-electron kinematics with a cross-section that has been adjusted for the effects of soft-photon emission (Eq. \ref{standardRC}).  In the hard regime, they are described by single-photon bremsstrahlung events.  The inclusion of both of these calculations allows the effects of photons of any energy to be considered.  The calculations have been formulated in the center-of-mass frame to take advantage of the many kinematic simplifications.

\subsection{Elastic Events with Soft-Photon Radiative Corrections}

Events with photons below the $\Delta E$ threshold are described with elastic kinematics and a cross-section that has been adjusted from Born as in Eq. (\ref{standardRC}).  The Born cross-section in the center-of-mass frame is given by

\begin{equation}
\frac{\mathrm{d}\sigma}{\mathrm{d}\Omega_3}{\bigg \vert}_\text{Born} = \frac{{\cal S}\langle|M|^2\rangle}{64\pi^2 \text{s}}
\label{borncs}
\end{equation}

\noindent with the tree-level matrix element for M\o ller scattering given by

\begin{align}
\langle|M|^2\rangle = 64\pi^2\alpha^2\Big[&\frac{m^4}{\text{t}^2}\Big(\frac{\text{s}^2+\text{u}^2}{2m^4} + 4\frac{\text{t}}{m^2}-4\Big)\nonumber \\
+&\frac{m^4}{\text{u}^2}\Big(\frac{\text{t}^2+\text{s}^2}{2m^4}+4\frac{\text{u}}{m^2}-4\Big) \nonumber \\
+&\frac{m^4}{\text{u}\text{t}}\Big(\frac{\text{s}}{m^2}-2\Big)\Big(\frac{\text{s}}{m^2}-6\Big)\Big].
\end{align}

\noindent Here, s, t, and u are the Mandelstam variables and $\Omega_3$ refers to the solid-angle of a particular final-state lepton.  The quantity ${\cal S}$ is a symmetry factor typically equal to $\prod_j 1/n_j!$ for each $n$ final-state identical particles of type $j$\footnote{For real experiments measuring M\o ller scattering, care must be taken to properly account for both final-state electrons.  When integrating over a non-trivial $\Omega_3$ region, the symmetry factor ${\cal S}$ may become a complicated function, especially for events with hard photons.}.  The matrix element for Bhabha scattering can easily be obtained from crossing symmetry by substituting $\text{s}\leftrightarrow\text{u}$.

Kaiser's derivation of the $\delta$ radiative correction terms is presented in \cite{softcorrections}.  To produce these corrections, the cross-section for soft-photon emission is first integrated over all photon directions and energies up to $\Delta E$.  The result of this is expressed as a correction to the Born cross section; it is, however, infrared-divergent.  An additional correction describing the interference between the tree-level and one-loop diagrams, however, contains an opposite infrared divergence \cite{softcorrections}.  Including both corrections thus produces a finite $\delta$ that can be used as in Eq. (\ref{standardRC}).

Equations (22) and (24) in \cite{softcorrections} provide the terms corresponding to soft-photon emission in M\o ller and Bhabha scattering, respectively.  While these terms contain the necessary cancellation of infrared divergences, they are incomplete because they do not describe the entirety of the effects from the one-loop diagrams.  As the text indicates, additional terms must be included to achieve a complete description \cite{softcorrections}.  This remaining part of the radiative correction is provided by summing the remaining finite loop-level interference terms and dividing them by the Born terms (i.e., second line of Kaiser's equation (2) divided by the first).  The expressions needed to compute this are printed in full for M\o ller scattering, but the corresponding Bhabha expressions can easily be obtained by the substitution $\text{s}\leftrightarrow\text{u}$.  The addition of these ($\Delta E$-independent) loop-level terms to the soft-photon expressions completes the description of the $\delta$ radiative correction factors for both M\o ller and Bhabha scattering.  We also note that we have included the terms containing both electronic and muonic vacuum polarization, although the latter is negligible at the energies we are considering.

One should note that as $\Delta E$ approaches zero, the soft-photon radiative corrections diverge to negative infinity.  This results from neglecting the effects of multiple soft-photon emission.  The effect of multiple soft photons can be taken into account to all orders by exponentiating the correction term ($1+\delta \rightarrow e^\delta$) \cite{exp}.  However, since we consider only single hard-photon bremsstrahlung, this would give the total cross-section an artificial dependence on $\Delta E$; as a result, the exponentiation is not used.  Our approach is self-consistent as long as $\Delta E$ is chosen to be large enough that the correction term remains small, but not so large that the soft-photon approximation becomes invalid.  Later in this paper we will examine some results with $\Delta E = 10^{-4}\sqrt{\text{s}}$.

We note that while we do not consider them, higher-order and multiple-photon effects may not be negligible when ${\cal O}$(0.1\%) absolute accuracy is desired.  In the case of DarkLight, the single-photon model is sufficiently precise, as we are largely interested in the noise created by the interaction of M\o ller electrons/photons with the detector elements.  For OLYMPUS, it is more important that the M\o ller and Bhabha processes be treated on equal footing, since the relevant quantity is the ratio of the cross-sections rather than the absolute value.   The framework used here is not easily scalable to include higher-order effects and multiple photons in a precise manner.   A different approach, such as a QED Parton Shower algorithm like that used in BabaYaga \cite{babayaga1,babayaga2}, is more suited to analyzing multiple-photon events; however, neither method is perfect and both do require some level of approximation.  

\subsection{Hard Bremsstrahlung Events}
\begin{figure}
\centering
\includegraphics[width=0.475\textwidth]{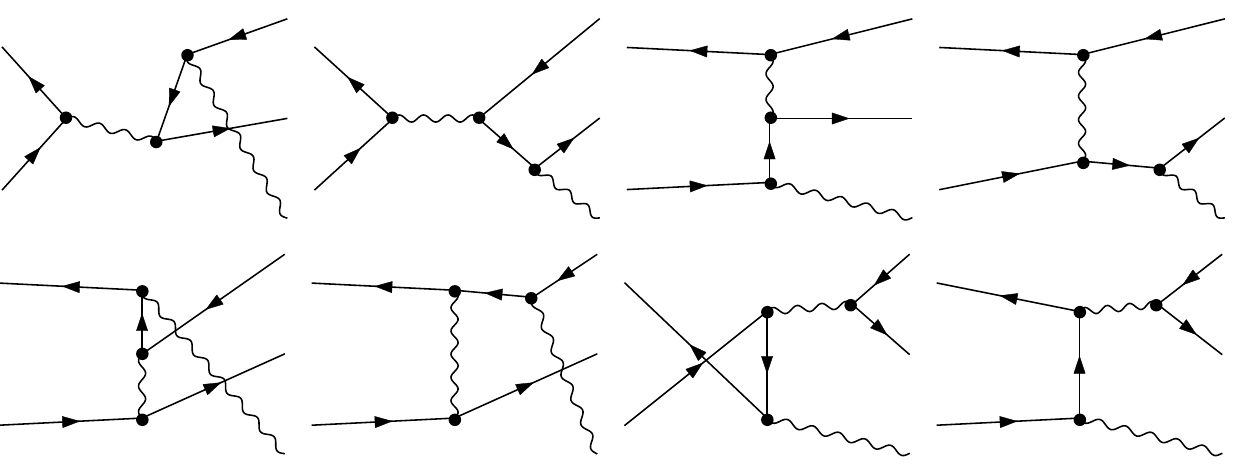}
\caption{Feynman diagrams for radiative Bhabha scattering}
\label{bhabhadiagrams}
\end{figure}

Events with photons having energy greater than $\Delta E$ are described by an exact tree-level single-photon bremsstrahlung calculation.  The spin-averaged matrix elements for $$e^{-}_1 + e^{-}_2 \rightarrow e^{-}_3 + e^{-}_4 + \gamma$$  and  $$e^+_1 + e^-_2 \rightarrow e^+_3 + e^-_4 + \gamma,$$ as diagrammed in Figs. \ref{mollerdiagrams} and \ref{bhabhadiagrams}, were calculated exactly using the Mathematica plugins \verb+FeynArts+ and \verb+FormCalc+ \cite{formcalc}.  No ultra-relativistic, soft-photon, or peaking approximations were made.

In formulating the center-of-mass phase-space parametrization for $2\rightarrow3$ body $e e \rightarrow ee\gamma$ scattering, we follow the approach of \cite{brem}.  Combined with the matrix elements, the bremsstrahlung cross-section is then given by:

\begin{equation}\frac{\mathrm{d}^5\sigma}{\mathrm{d}E_\gamma\ \mathrm{d}\Omega_\gamma\ \mathrm{d}\Omega_3}=\frac{\cal S}{32m^2(2\pi)^5}\frac{E_\gamma }{2Ep{\vartheta}}\sum_\nu p_{3\nu}^2\langle|M|^2\rangle
\label{cs}
\end{equation}

\noindent with

\begin{equation}{\vartheta} = \frac{1}{m}\sqrt{4E^2(E-E_\gamma)^2/m^2-(2E-E_\gamma)^2+E_\gamma^2\cos^2\alpha},
\label{rc}
\end{equation}

\noindent where $\alpha$ is the angle between lepton 3 and the photon, $E$ and $p$ are the center-of-mass frame energy/momentum of either initial-state particle, and $m$ is the electron mass.

The energy of lepton 3 is then given by \cite{brem}:

\begin{equation}E_{3} = \frac{2E(E-E_\gamma)(2E-E_\gamma)\mp m^2 E_\gamma {\vartheta}\cos\alpha }{(2E-E_\gamma)^2-E_\gamma^2\cos^2\alpha}
\label{eq1}.
\end{equation}

\noindent If the photon energy is below

\begin{equation}E_{\gamma_0} = 2E(E-m)/(2E-m),
\label{k0}
\end{equation}

\noindent then only the upper sign in Eq. (\ref{eq1}) is allowed.  If it is above $E_{\gamma_0}$, both are allowed, and there is an additional constraint that

\begin{equation}\cos\alpha < -\frac{1}{E_\gamma}\sqrt{(2E-E_\gamma)^2-4E^2(E-E_\gamma)^2/m^2}.
\label{cosa}
\end{equation}

\noindent The summation in Eq. (\ref{cs}) indicates that both possible values, i.e., both signs in Eq. (\ref{eq1}) should be included in the case that $E_\gamma > E_{\gamma_0}$ where both are valid.  This cutoff, $E_{\gamma_0}$, is purely an artifact of this choice of variables; however, these variables are necessary in order to properly match the soft-photon and hard-photon parts of the cross-section, by defining hard photons as those with $E_\gamma > \Delta E$.  We also note that the highest possible photon energy is equal to 

\begin{equation}E_{\gamma_{\text{max}}} = p^2/E = E - m^2/E,
\label{maxE}
\end{equation}

\noindent which occurs when the two outgoing leptons are emitted collinearly opposite the photon, each carrying half its momentum.  

\begin{figure}
\centering
\subfigure[M\o ller Scattering]{
\includegraphics[width=0.5\textwidth]{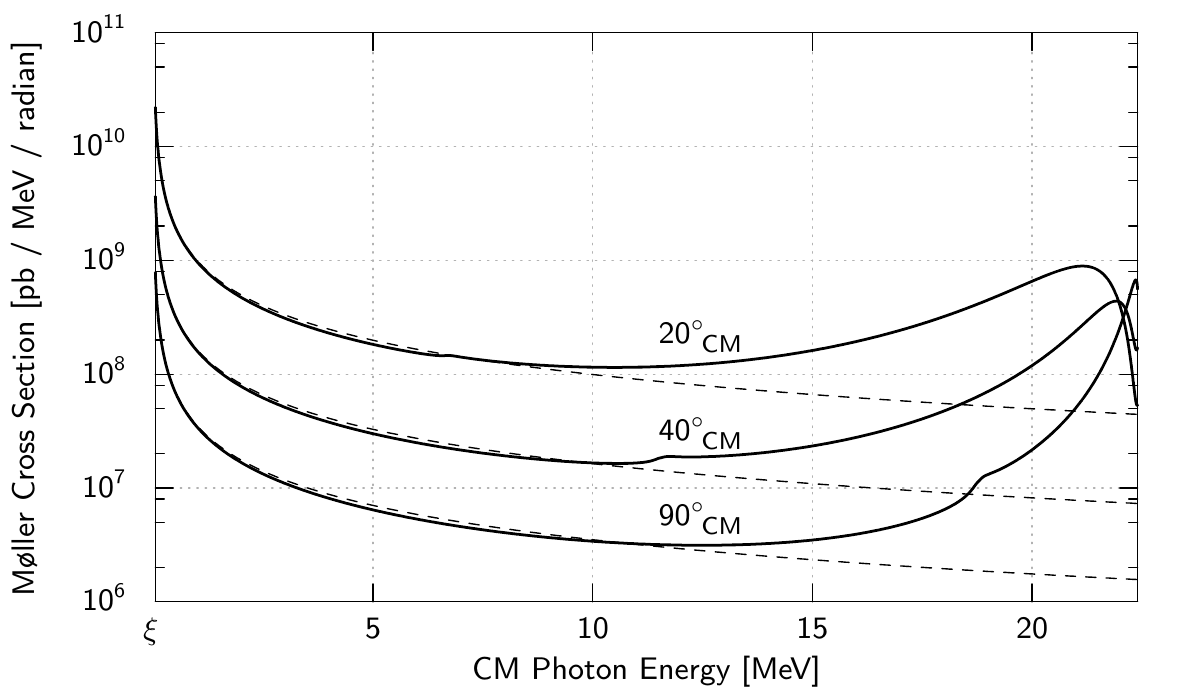}}

\subfigure[Bhabha Scattering]{
\includegraphics[width=0.5\textwidth]{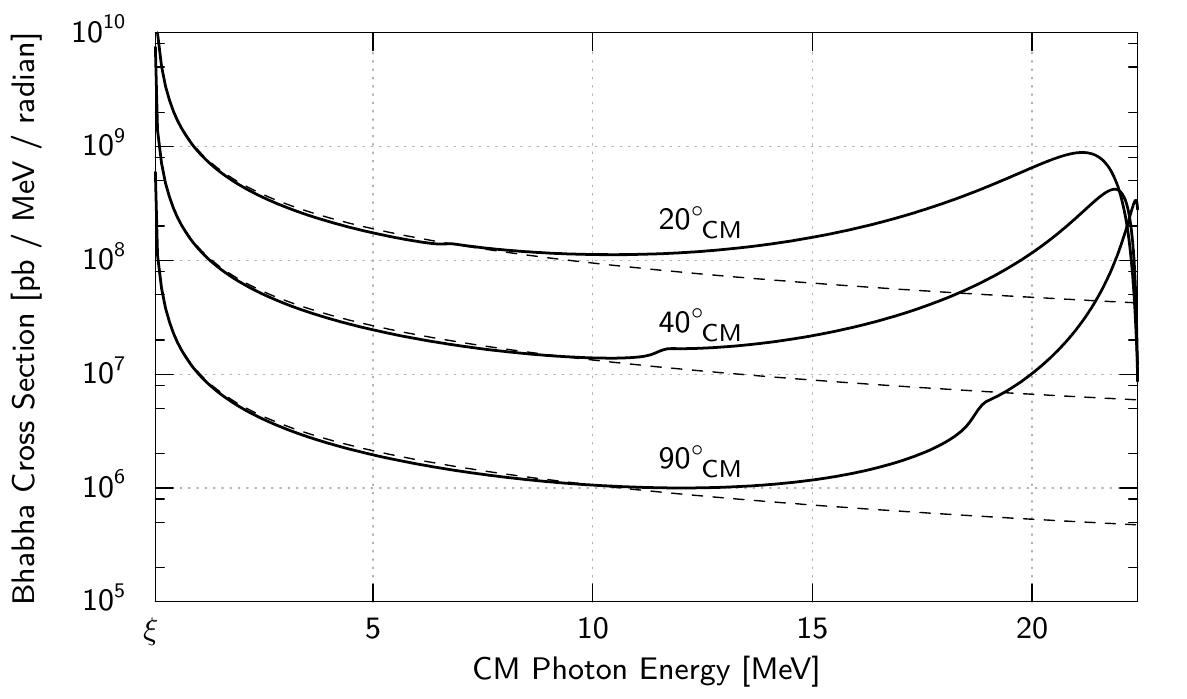}}
\caption{Cross-sections for hard bremsstrahlung (solid lines) compared with soft-photon corrections (dashed lines \cite{softcorrections}) at various center-of-mass frame lepton angles for M\o ller and Bhabha Scattering, for the range $\xi < E_\gamma < E_{\gamma_0}$.}
\label{verification}
\end{figure}

\begin{figure}
\centering
\subfigure[M\o ller Scattering]{
\includegraphics[width=0.5\textwidth]{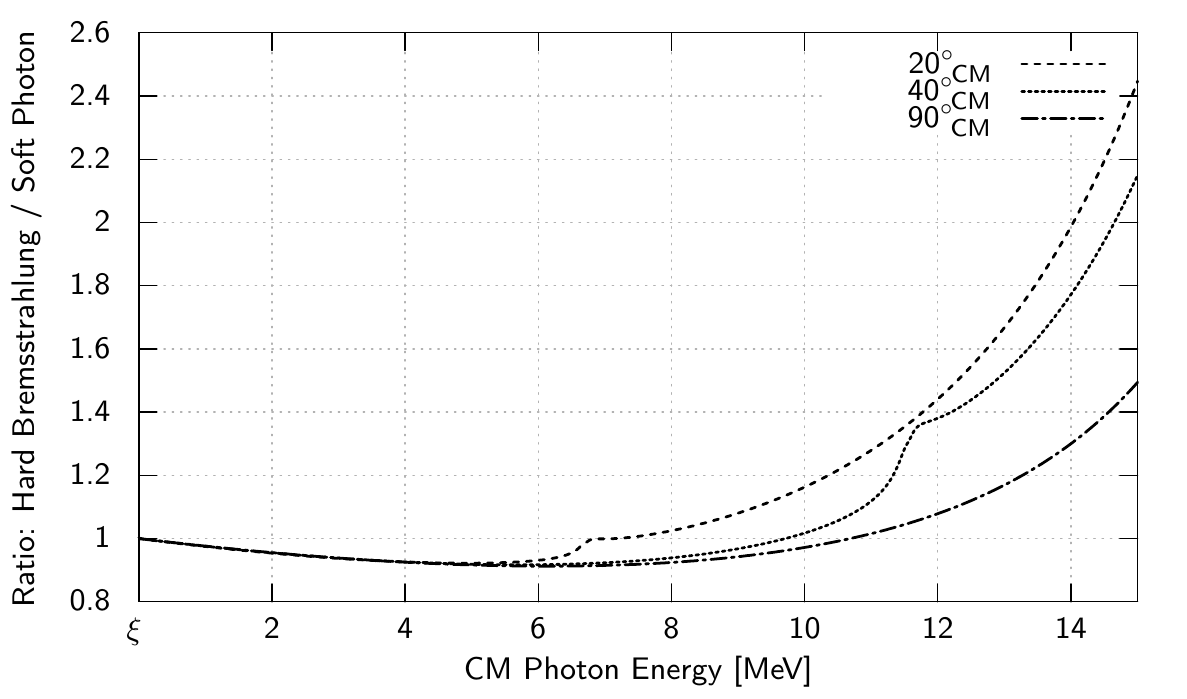}}

\subfigure[Bhabha Scattering]{
\includegraphics[width=0.5\textwidth]{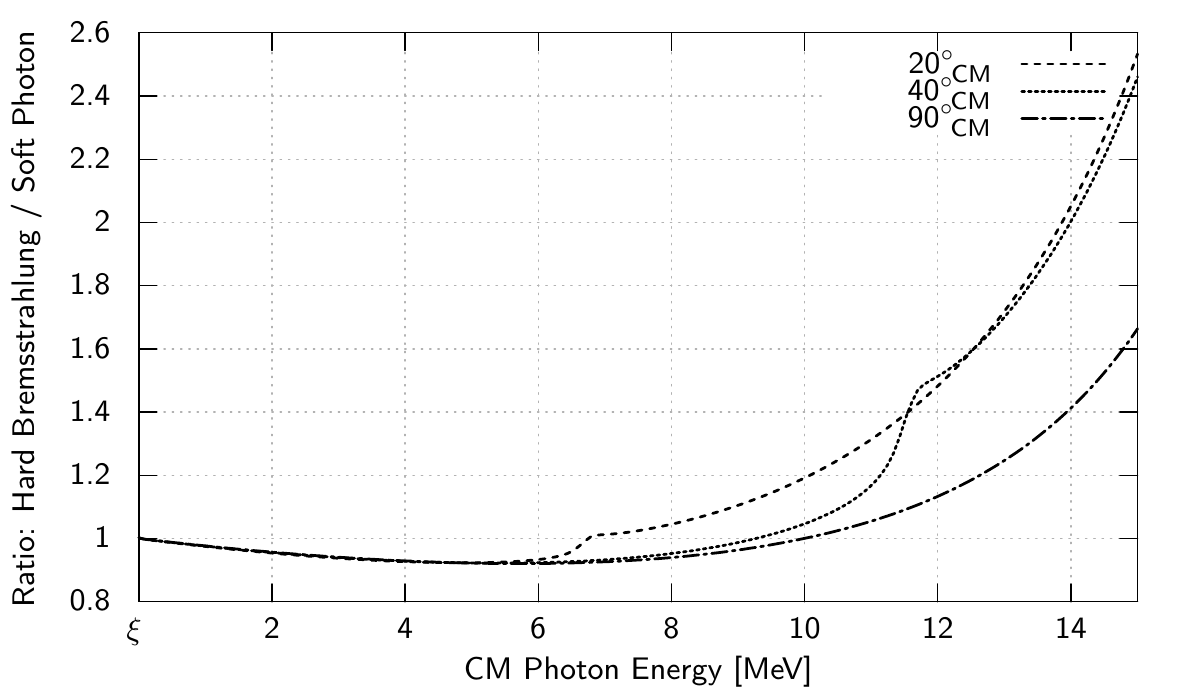}}
\caption{Ratio of hard bremsstrahlung (Eq. \ref{hardgamma}) to the soft-photon corrections (Eq. \ref{softgamma}). The agreement as $E_\gamma\rightarrow\xi$ indicates the bremsstrahlung behaves as expected.  Deviations from unity are expected as the soft-photon approximation breaks down.}
\label{verification2}
\end{figure}

\begin{figure}
\centering
\subfigure[M\o ller Scattering]{
\includegraphics[width=0.5\textwidth]{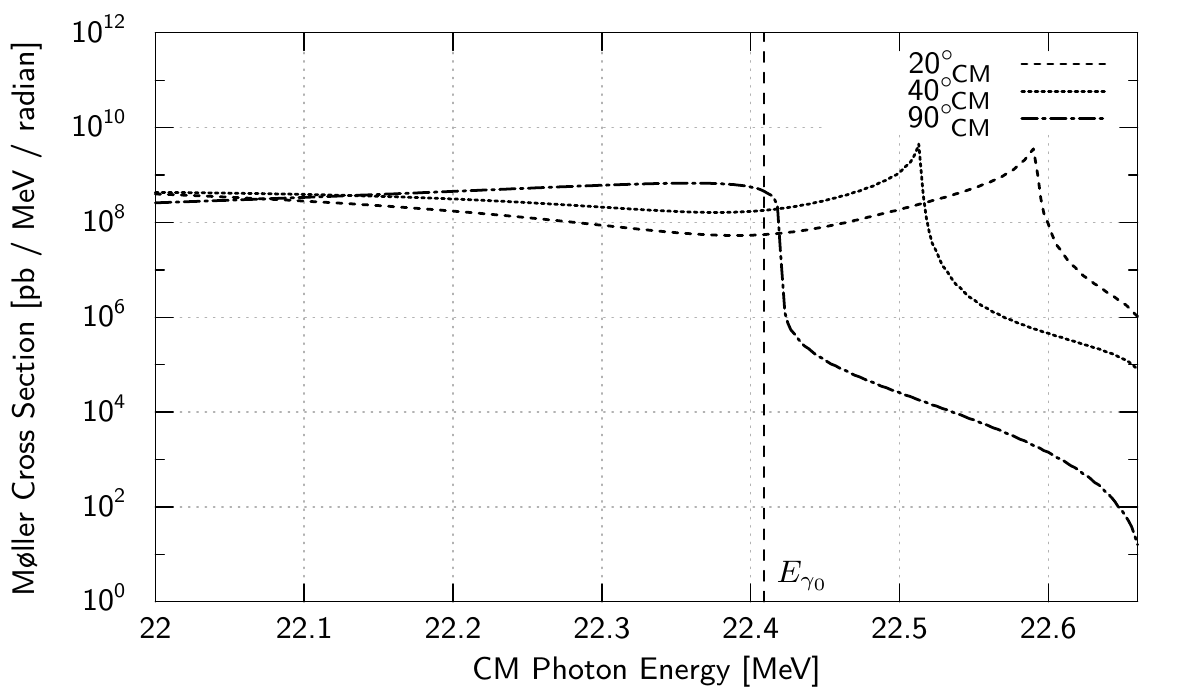}}

\subfigure[Bhabha Scattering]{
\includegraphics[width=0.5\textwidth]{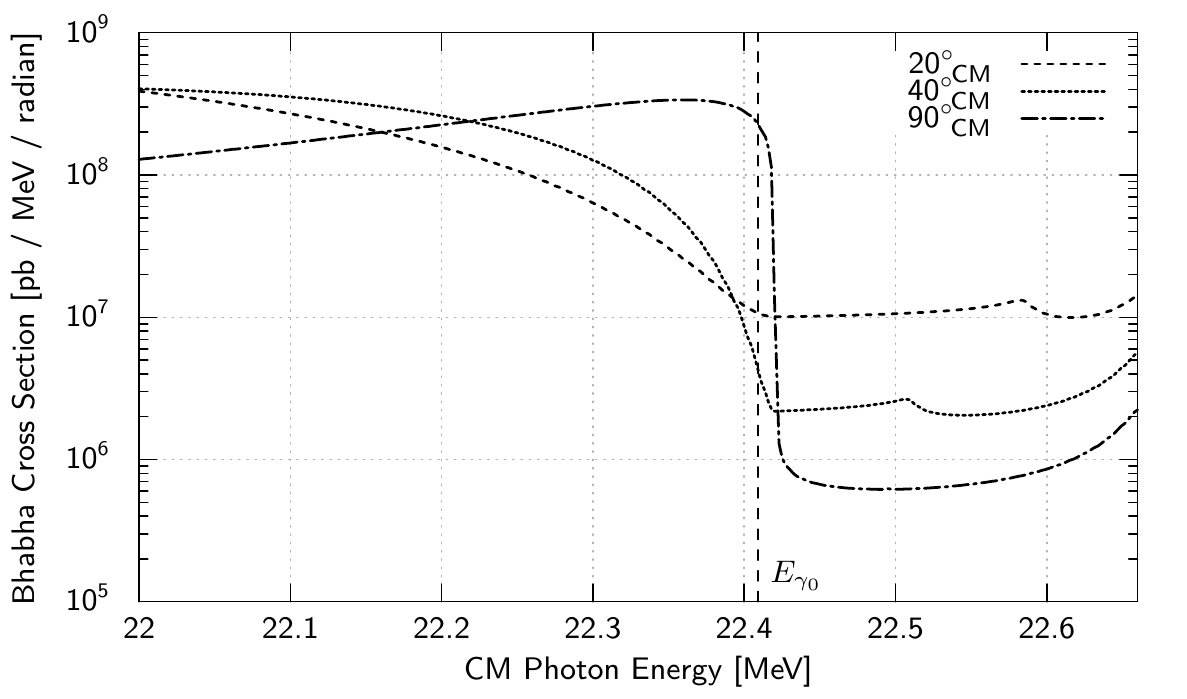}}
\caption{Bremsstrahlung cross-section at various center-of-mass frame lepton angles for M\o ller and Bhabha Scattering, plotted at the highest-allowable photon energies.}
\label{verification3}
\end{figure}

\begin{figure}
\centering
\includegraphics[width=0.5\textwidth]{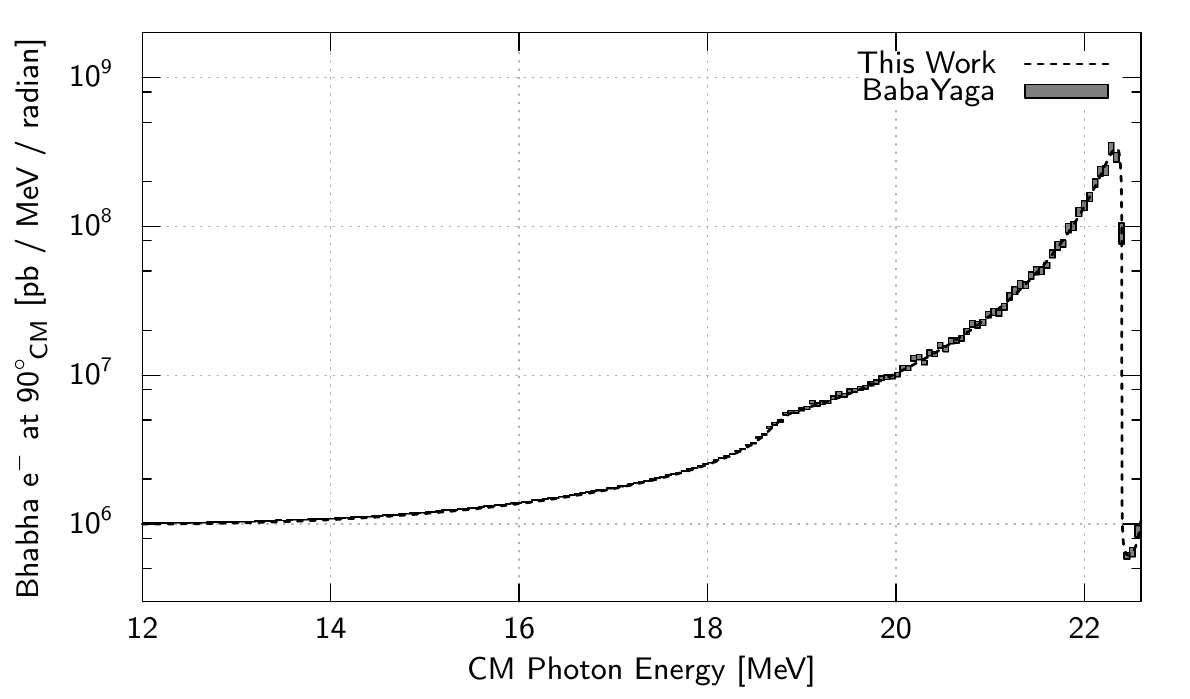}
\caption{This work (dashed line) compared with BabaYaga@NLO at order alpha (boxes), for detecting an electron at 90$^\circ$ in the CM frame, as a function of photon energy.  Box height (not visible at lower photon energies) indicates statistical Monte Carlo error.}
\label{compare}
\end{figure}

\section{Discussion of Results}

In the following section, we present some results at a center-of-mass energy of $\sqrt{\text{s}} = 45.3$ MeV, corresponding to OLYMPUS kinematics of a 2.01 GeV beam incident on a fixed target.
These results have been calculated with $\Delta E = 10^{-4}\sqrt{\text{s}} \approx 4.5$ keV; we will refer to this particular cut-off value as $\xi$.
In Fig. \ref{verification}, a comparison between the hard-photon bremsstrahlung cross-section and the soft-photon-corrected cross-section is presented at three specific lepton angles for $\xi < E_\gamma < E_{\gamma_0}$.  The bremsstrahlung cross-section has been numerically integrated over all photon directions, and is plotted as a function of photon energy.  The soft-photon-corrected cross section has been differentiated with respect to $\Delta E$ to obtain a cross-section as a function of photon energy.  This formulation produces two quantities that can be directly compared:

\begin{align}
\text{Soft:}\ &\frac{\mathrm{d}^3\sigma}{\mathrm{d}\Omega_3\ \mathrm{d}E_\gamma} = \frac{\mathrm{d}}{\mathrm{d}\Delta E}\big\{\delta(\Omega_3,\Delta E)\big\} \times \frac{\mathrm{d}\sigma}{\mathrm{d}\Omega_3}{\bigg \vert}_\text{Born} \label{softgamma}\\
\text{Hard:}\ &\frac{\mathrm{d}^3\sigma}{\mathrm{d}\Omega_3\ \mathrm{d}E_\gamma} = \int_{4\pi} \frac{\mathrm{d}^5\sigma}{\mathrm{d}\Omega_3\ \mathrm{d}E_\gamma\ \mathrm{d}\Omega_{\gamma}}\ \mathrm{d}\Omega_\gamma.\label{hardgamma}
\end{align}

\noindent These M\o ller (Bhabha) cross-sections represent the probability for detecting an electron (positron) at the specified angle as a function of the energy of the emitted photon.  At low photon energies, the close agreement is a validation of our code and is a reflection that the calculations properly reduce to existing soft-photon calculations at $E_\gamma = \xi$.  Figure \ref{verification2} shows a ratio of these quantities; here, the agreement can be clearly seen by the ratio becoming unity as $E_\gamma \rightarrow \xi$.

The soft-photon cross-section (Eq. \ref{softgamma}) has been plotted to photon energies that are clearly outside its range of validity in order to demonstrate its limitations.  At these higher photon energies, a relative rise of the hard-photon bremsstrahlung cross-section is seen, corresponding to an increase of the cross-section resulting from initial-state radiation.  Figure \ref{verification3} shows the hard cross-sections plotted at the highest photon energies.  We also note that the M\o ller cross-sections presented in Figs. \ref{verification}(a) and \ref{verification3}(a) are that for detecting \emph{any} electron, and may exceed other formulations by a factor of two.

In many of these plots, features such as kinks and cusps are visible, especially in the region where $E_\gamma > E_{\gamma_0}$.  However, we note that in this scenario with a very high-energy photon, final-state leptons are emitted nearly collinearly (in the CM frame), and in this region the single-photon bremsstrahlung model may break down.  Contributions from multiple-photon exchange/emission, final-state interactions, and atomic effects may become important in this regime.  We are able to reproduce these features for the M\o ller case with the matrix element presented in \cite{brem}.  In addition, excellent agreement with the widely-used code BabaYaga@NLO \cite{babayaga2} (which includes the electron mass) is observed when run at order alpha (NLO).  Figure \ref{compare} shows a comparison between our work and BabaYaga, in which the interesting features line up precisely.  There is an approximately 1\% deviation between our work and BabaYaga in the mid-photon-energy region ($\sim$12 MeV), but this is only at the lowest point of the cross-section, and likewise it contributes negligibly to the total cross-section.  This may result from approaching the same physics with contrasting methods.


An upcoming opportunity to verify the results of this paper is the Phase 1 run of the DarkLight experiment.  A primary goal of Phase 1 is to measure various Standard Model processes at 100 MeV, including elastic electron-proton scattering and radiative M\o ller scattering.  A dedicated experimental apparatus is being realized to measure radiative M\o ller scattering.  It is envisioned that data will be acquired in this run to precisely verify the calculations described in this paper.

\section{Summary}

A formulation of the next-to-leading-order radiative corrections to M\o ller and Bhabha scattering has been prepared, including hard-photon effects and avoiding ultra-relativistic approximations.  It realizes a treatment of events with single hard-photon emission, as well as the effects of soft-photon emission from events well-described by elastic kinematics.  Information about behavior at a large range of photon energies is thus provided in a way that can easily be incorporated into a Monte Carlo simulation via a newly-developed event generator.  It is well-suited for electron and positron beam experiments, such as DarkLight and OLYMPUS, as a basis for simulations to study backgrounds as well as to precisely measure luminosity.  A direct validation of the calculation with data is being investigated for the upcoming Phase 1 DarkLight experiment at Jefferson Lab.

\begin{acknowledgements}
We are grateful to Axel Schmidt, Colton D. O'Connor, and Jan C. Bernauer for their help and careful guidance in preparing this work.  We would also like to thank Peter Blunden and T. William Donnely for their generous feedback and very helpful discussions.  This research is supported by the U.S. Department of Energy Office of Nuclear Physics under grant number DE-FG02-94ER40818, a Robert Lourie Graduate Fellowship of the MIT Department of Physics, and a U.S. Department of Energy National Nuclear Security Administration Stewardship Science Graduate Fellowship (provided under grant number DE-NA0002135).
\end{acknowledgements}
\bibliography{RadMoller}

\end{document}